\input{aipcheck}

\documentclass[
    ,final            
  ]
  {aipproc}

\usepackage{amsmath}
\usepackage{amssymb}

\layoutstyle{6x9}

\newcommand{\etal}{{\em et al.}}                

\newcommand{\beq}{\begin{equation}}
\newcommand{\eeq}{\end{equation}}

\begin{document}

\title{Symmetry Energy from Systematic\\ of Isobaric Analog States}


\author{Pawel Danielewicz}{
  address={National Superconducting Cyclotron Laboratory and
Department of Physics and Astronomy, Michigan State University,
East Lansing, Michigan 48824, USA}
}

\author{Jenny Lee}{
  address={RIKEN Nishina Center for Accelerator-Based Science\\
Wako, Saitama 351-0198, Japan}
}

\begin{abstract}
Excitation energies to isobaric states, that are analogs of ground states, are dominated by contributions from the symmetry energy. This opens up a possibility of investigating the symmetry energy on nucleus-by-nucleus basis. Upon correcting energies of measured nuclear levels for shell and pairing effects, we find that the lowest energies for a given isospin rise in proportion to the square of isospin, allowing for an interpretation of the coefficient of proportionality in terms of a symmetry coefficient for a given nucleus. In the (A,Z) regions where there are enough data, we demonstrate a Z-independence of that coefficient. We further concentrate on the A-dependence of the coefficient, in order to learn about the density dependence of symmetry energy in uniform matter, given the changes of the density in the surface region. In parallel to the analysis of data, we carry out an analysis of the coefficient for nuclei calculated within the Skyrme-Hartree-Fock (SHF) approach, with known symmetry energy for uniform matter. While the data from isobaric analog states suggest a simple interpretation for the A-dependent symmetry coefficient, in terms of the surface and volume symmetry coefficients, the SHF results point to a more complicated situation within the isovector sector than in the isoscalar, with much stronger curvature effects in the first. We exploit the SHF results in estimating the curvature contributions to the symmetry coefficient. That assessment is hampered by instabilities of common Skyrme parameterizations of nuclear interactions.
\end{abstract}

\maketitle


\section{Symmetry Energy}

Symmetry energy describes how bulk nuclear energy evolves with changes in neutron-proton asymmetry.  It is naturally of great interest in the context of extrapolations from nuclear systems observed under laboratory conditions to neutron matter.  Moreover, because of different contributions of the two- and few-nucleon interactions to neutron-proton symmetric and to neutron-dominated systems, it is of interest in sorting out the impact of those interactions on nuclear properties.

Energy of a nucleus with $N$ neutrons and $Z$ protons, $N+Z = A$, may be represented in the form
\beq
E(N,Z) = E_0 (A) + E_1 + E_C + E_\text{mic} \, .
\eeq
Here, $E_0$ and $E_1$ stand for bulk nuclear contributions to the energy, with $E_0$ -- the energy of a symmetric system and $E_1$ -- the symmetry energy.  Further, $E_C$ is Coulomb energy and $E_\text{mic}$ represents microscopic corrections including shell effects and pairing.  Charge symmetry implies that the symmetry energy should be a quadratic function of the asymmetry $(N-Z)$:
\beq
\label{eq:E1}
E_1 = \frac{a_a(A)}{A} \, (N - Z)^2 \, .
\eeq
Here, $a_a(A)$ is a generalized mass-dependent (a)symmetry coefficient~\cite{Danielewicz200936}.  Charge symmetry considerations for uniform nuclear matter, with neutron and proton densities of $\rho_n$ and $\rho_p$, respectively, $\rho = \rho_n + \rho_p$, and Coulomb interactions switched off, similarly yield:
\beq
\frac{E}{A} (\rho_n , \rho_p) = \frac{E_0}{A} (\rho) + \frac{E_1}{A} = \frac{E_0}{A} (\rho)
+ S(\rho) \, \Big( \frac{\rho_n - \rho_p}{\rho} \Big)^2 \, .
\eeq
In the literature, both $E_1$ and the coefficient $S(\rho)$ are referred to as symmetry energy.  The density-dependent symmetry energy $S$ is usually expanded around normal density~$\rho_0$:
\beq
\label{eq:Srho}
S(\rho) = a_a^V + \frac{L}{3} \, \frac{\rho-\rho_0}{\rho_0} + \ldots \, .
\eeq
On account of energy of symmetric matter $E_0$ minimizing at $\rho_0$, at nuclear densities, pressure in neutron matter with energy $E=E_1+S$, sustaining neutron stars, tends to be dominated by the symmetry energy.  Naturally that pressure is strongly sensitive to the value of $L$ in \eqref{eq:Srho}.
We represent the value of the symmetry energy at $\rho_0$ with a symmetry coefficient in~\eqref{eq:Srho}, $S(\rho_0) \equiv a_a^V$, because the generalized symmetry coefficient of Eq.~\eqref{eq:E1} should reduce to $S(\rho_0)$ in the limit of $A \rightarrow \infty$.

Changes in nuclear density across the surface, interplaying with the density dependence of the symmetry energy in uniform matter, should generally make the symmetry coefficient $a_a$ dependent on mass.  In fact, a simple consideration of the effects of volume-surface competition on symmetry energy \cite{Danielewicz200936,Danielewicz:2003dd} produces the result
\beq
\label{eq:aaA}
\frac{1}{a_a(A)} = \frac{1}{a_a^V} + \frac{A^{-1/3}}{a_a^S} \, .
\eeq
As $A \rightarrow \infty$, the coefficient $a_a(A)$ reduces here to $a_a^V$; $a_a^S$ is the surface symmetry coefficient related to the $\rho$-dependence of $S$ and to $L$ in particular~\cite{Danielewicz200936}. Following the above, establishing the mass dependence of the coefficient could provide information on $S(\rho)$ and constrain pressures in neutron stars.

Unfortunately, when trying to deduce an $A$-dependence of the symmetry coefficient, by fitting a bulk-energy formula with \eqref{eq:E1} to observed nuclear masses, the results turn out to be ambiguous~\cite{Danielewicz:2003dd}, even for such a simple parameterization as in Eq.~\eqref{eq:aaA}.  This is because the effects of the symmetry energy compete with the effects of Coulomb energy and even with terms that contain no asymmetry, due to the mass-asymmetry correlation induced by the line of stability.  Fortunately, extending the bulk-energy formula, to reach consistency with the charge invariance of nuclear interactions, allows for an analysis of the symmetry energy on a nucleus-by-nucleus basis, independently of Coulomb energy and energy of symmetric matter.

\section{Analysis of Isobaric Analog States}

Extension of the bulk-energy formula relies on the observation that the symmetry energy~\eqref{eq:E1} may be rewritten in terms of isospin as
\beq
E_1 = \frac{4 \, a_a(A)}{A} \, T_z^2 \, .
\eeq
However, nuclear contributions to the energy should be scalars under rotations in isospin space~\cite{BlattWeisskopf}, with $z$ and transverse isospin components treated democratically.  Thus, the symmetry energy should be rather written as
\beq
E_1 = \frac{4 \, a_a(A)}{A} \, {\pmb T}^2 = \frac{4 \, a_a(A)}{A} \, \big( T_z^2 + {\pmb T}_\perp^2    \big)   \,  = \frac{4 \, a_a(A)}{A} \, T (T+1) \, .
\eeq
With this extension for the symmetry energy, one can consider states of lowest energy in a given nucleus, for a fixed isospin~$T$, rather than absolute ground states that normally have $T = |T_z|$.  Those excited states end up being isobaric analog states (IAS) of ground states of other nuclei within the same isobaric chain.  Regarding the bulk contributions to the energy, the $N=Z$ portion is the same for those states as for the ground state of the specific nucleus and the Coulomb contribution is the same as well.  The excitation energy to such a state should then reflect change in the symmetry energy associated with the change in isospin
\beq
\label{eq:EIAS}
E_\text{IAS}^* = \frac{4 \, a_a(A)}{A} \, \Delta {\pmb T}^2 + \Delta E_\text{mic} =  \frac{4 \, a_a(A)}{A} \, \big( {\pmb T}_\perp^2 - |T_z| \big) + \Delta E_\text{mic} \, ,
\eeq
where the isospin in the last term pertains to IAS and where we used $T = |T_z|$ for the ground state.

\begin{figure}
\centerline{\includegraphics[width=.75\linewidth]{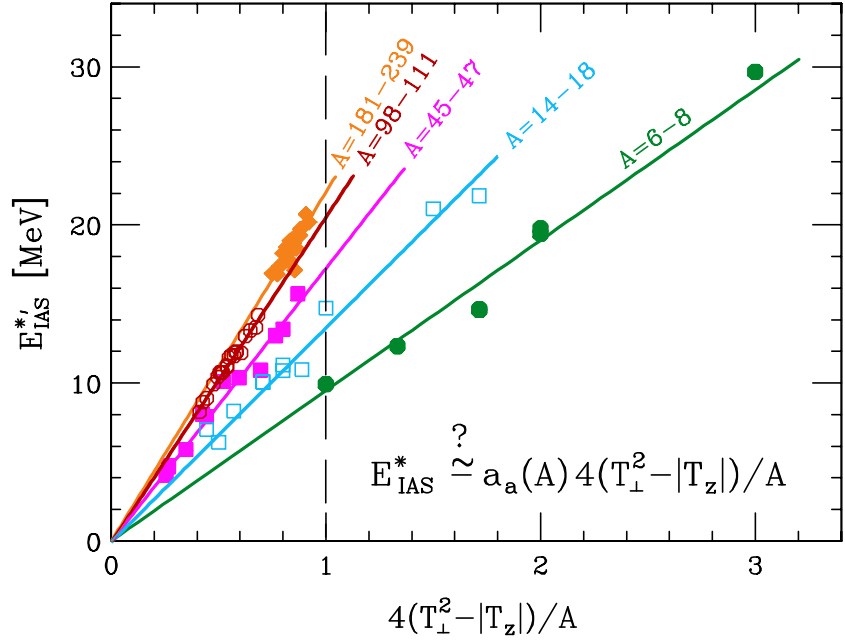}}
\caption{Excitation energies to the ground-state IAS, corrected for microscopic effects~\cite{Koura:2005}, plotted vs change in $T(T+1)$ scaled with~$A$, represented by different symbols for different indicated intervals in the mass number $A$.  The solid lines represent linear least-square fits to the represented energies.  The fits are forced to pass through the coordinate origin.  The dashed vertical line helps to read off the values of $a_a(A)$ for the specific $A$-intervals, from intersection of the fitted lines with the vertical.
}
\label{fig:EIASTTA}
\end{figure}

To test whether Eq.~\eqref{eq:EIAS} could be valid, we use the measured energies of IAS, primarily from the compilation \cite{Antony:1997}, and construct excitation energies corrected for microscopic and deformation effects, following the results of~\cite{Koura:2005}.  For selected narrow intervals in mass, we plot those energies in Fig.~\ref{fig:EIASTTA}, vs scaled difference in isospin squared, between the IAS and ground state.  If the generalized symmetry coefficient varies slowly with mass, the excitation energies should line up, according to Eq.~\eqref{eq:EIAS}, along the lines passing through the origin, with a slope given by the symmetry coefficient.  It is seen in Fig.~\ref{fig:EIASTTA} that the excitation energies indeed line up in such a manner, with line slopes increasing as the mass number increases.

\begin{figure}
\centerline{\includegraphics[width=.85\linewidth]{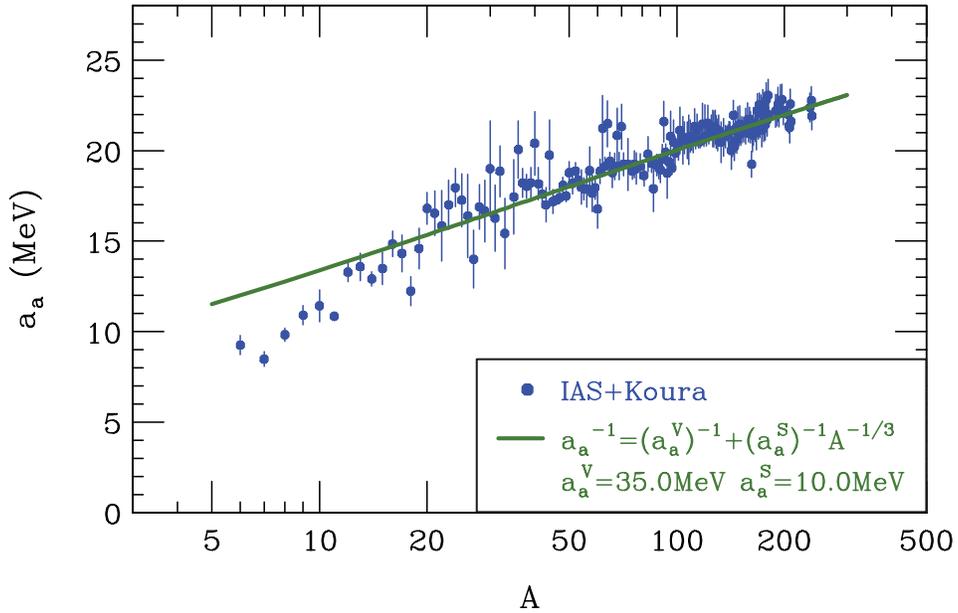}}
\caption{Generalized mass-dependent asymmetry coefficient $a_a(A)$ extracted from excitation energies to ground-state IAS within individual isobaric chains $A$, when applying shell corrections by Koura \etal
The line represents a fit at $A \ge 22$ assuming a combination of volume and surface symmetry terms.
}
\label{fig:asyme}
\end{figure}

The generalized symmetry coefficients, obtained by fitting the IAS excitation spectra at a given~$A$, are next shown as a function of~$A$ in Fig.~\ref{fig:asyme}.  The errors in the figure are estimated assuming residual unsubtracted microscopic contributions to the input energies, of an r.m.s.\ magnitude of $\delta_\text{rms} = 0.5 \, \text{MeV}$.  A significant variation with mass is observed for coefficient values.  The typical expectation of $a_a \gtrsim 20 \, \text{MeV}$ is met for mass numbers $A > 100$.  However, for $A < 10$ even as low values as $a_a \sim 10 \, \text{MeV}$ are reached.    It is apparent that for masses $A \gtrsim 20$, the dependence of the coefficient on mass is well described by the volume-surface competition formula of Eq.~\eqref{eq:aaA}.  Depreciation of coefficient values with mass, at lower mass values, is faster than predicted by~\eqref{eq:aaA}.

\begin{figure}
\centerline{\includegraphics[width=.60\linewidth]{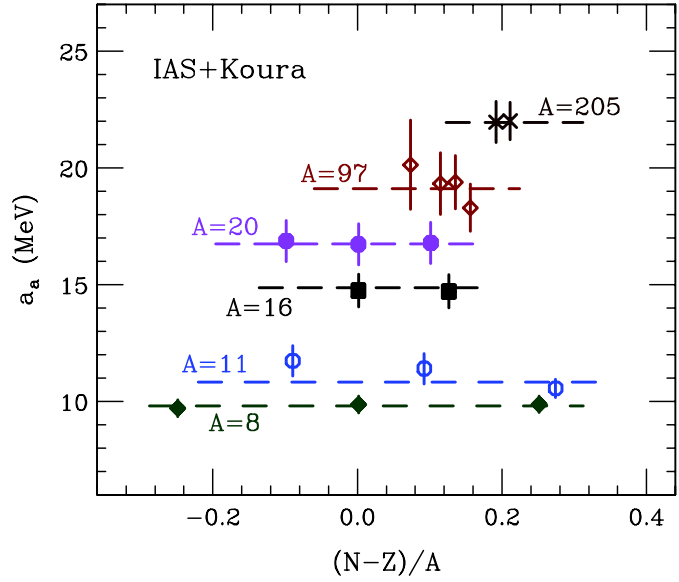}}
\caption{Generalized symmetry coefficient for individual nuclei, within selected isobar chains characterized by mass numbers~$A$, presented here as a function of asymmetry.  The horizontal dash lines represent the coefficient values for the isobaric chains as a whole.
}
\label{fig:asymz}
\end{figure}

We further test whether there is evidence for a $Z$-dependence of the generalized symmetry coefficient, in the context of claims made in Refs.~\cite{Janecke200323,Janecke2007317,PhysRevC.81.067302}.  Exemplary values of coefficients for different members of isobaric chains, with emphasis on the span of masses and coefficient values, as well as the number of members in a chain that could be analyzed, are shown in Fig.~\ref{fig:asymz}.  We find no evidence for the variation of coefficients with~$Z$.  

\section{From IAS to Infinite Matter}

Given the success of the fit with \eqref{eq:aaA}, to the results from IAS displayed in Fig.~\ref{fig:asyme}, it would be natural to proceed directly to the conclusions on properties of infinite matter~\cite{PhysRevLett.102.122701}.  Thus, the coefficient $a_a^V \simeq 35.0 \, \text{MeV}$ could be identified with the value of the symmetry energy at normal density $S(\rho_0)$.  Moreover, the surface symmetry coefficient $a_a^S \simeq 10.0 \, \text{MeV}$ could be identified with that for the half-infinite nuclear matter \cite{Danielewicz200936}.  From correlation between $a_a^S$ and slope~$L$ of symmetry energy, within the calculations of half-infinite matter~\cite{Danielewicz200936}, one could then deduce $L \sim 95 \, \text{MeV}$.  A more cautious strategy involves testing of the procedure within a semi-realistic description of nuclear properties such as Skyrme-Hartree-Fock (SHF) \cite{reinhard:014309,Danielewicz:2012}.

The complication with SHF is in the fact that the approach violates charge invariance, meaning that the procedure such as for the data cannot be applied.  Another complication is in the lack of shell corrections established in such a manner as for the data.  Correspondingly a different way of arriving at generalized asymmetry coefficients needs to be found \cite{Danielewicz:2012}.  When the coefficients are found, it is observed that for virtually every Skyrme interaction the approach to the limit of clear  surface-volume separation is much slower for the symmetry energy than for the energy of symmetric matter.  In consequence, for nuclei occurring in nature, the coefficients in Eq.~\eqref{eq:aaA} turn out to be effective rather than representing nuclear matter.  Interestingly, for many Skyrme parameterizations in the literature, the limit of Eq.~\eqref{eq:aaA} with coefficients truly represented nuclear matter is actually never reached, even for nuclei calculated~\cite{reinhard:014309} using $A \sim (100 \,000 - 1 \, 000 \, 000)$, while switching off Coulomb interactions.  This is because of instabilities \cite{Lesinski:2006cu} for many of the interactions, particularly isovector in nature, that come into play in large systems.

\section{Conclusions}

Symmetry coefficients can be obtained for individual nuclei, exploiting charge invariance of nuclear interactions.  The coefficients rise with mass number within the range of  $\sim (10 - 23) \, \text{MeV}$.  For $A \gtrsim 20$, the coefficients are well described with a surface-volume formula $a_a^{-1}(A) \simeq (a_a^V)^{-1} + (a_a^S)^{-1} \, A^{-1/3}$, where $a_a^V \simeq 35.0 \, \text{MeV}$ and $a_a^S \simeq 10.0 \, \text{MeV}$.  Analysis done with spherical SHF calculations suggests that the volume and surface coefficients from the fit are effective in nature rather than representing the infinite matter.  Assessment of the properties of the latter, using IAS results, is still in progress.


\begin{theacknowledgments}

This work was supported by the U.S.\ National Science Foundation under Grants PHY-0800026 and PHY-1068571 and by JUSTIPEN under U.S.\ Department of Energy Grant DEFG02-06ER41407.

\end{theacknowledgments}



\bibliographystyle{aipproc}   


\bibliography{coqu11}

\begin{thebibliography}{12}
\expandafter\ifx\csname natexlab\endcsname\relax\def\natexlab#1{#1}\fi
\providecommand{\enquote}[1]{``#1''}
\expandafter\ifx\csname url\endcsname\relax
  \def\url#1{\texttt{#1}}\fi
\expandafter\ifx\csname urlprefix\endcsname\relax\def\urlprefix{URL }\fi
\providecommand{\eprint}[2][]{\url{#2}}

\bibitem[Danielewicz and Lee(2009)]{Danielewicz200936}
P.~Danielewicz, and J.~Lee, \emph{Nucl. Phys. A} \textbf{818}, 36 -- 96 (2009).

\bibitem[Danielewicz(2003)]{Danielewicz:2003dd}
P.~Danielewicz, \emph{Nucl. Phys.} \textbf{A727}, 233--268 (2003),
  \eprint{nucl-th/0301050}.

\bibitem[Blatt and Weisskopf(1952)]{BlattWeisskopf}
J.~M. Blatt, and V.~I. Weisskopf, \emph{{Theoretical Nuclear Physics}}, John
  Wiley and Sons, New York, 1952.

\bibitem[Antony et~al.(1997)]{Antony:1997}
M.~S. Antony, A.~Pape, and J.~Britz, \emph{At. Data Nucl. Data Tables}
  \textbf{66}, 1--63 (1997).

\bibitem[Koura et~al.(2005)]{Koura:2005}
H.~Koura, T.~Uno, T.~Tachibana, and M.~Yamada, \emph{Prog. Theor. Phys.}
  \textbf{113}, 305--325 (2005).

\bibitem[Janecke et~al.(2003)]{Janecke200323}
J.~Janecke, T.~W. O'Donnell, and V.~I. Goldanskii, \emph{Nucl. Phys. A}
  \textbf{728}, 23 -- 51 (2003), ISSN 0375-9474,
  \urlprefix\url{http://www.sciencedirect.com/science/article/pii/S03759474030%
16865}.

\bibitem[Janecke and O'Donnell(2007)]{Janecke2007317}
J.~Janecke, and T.~O'Donnell, \emph{Nucl. Phys. A} \textbf{781}, 317 -- 341
  (2007), ISSN 0375-9474.

\bibitem[Wang and Liu(2010)]{PhysRevC.81.067302}
N.~Wang, and M.~Liu, \emph{Phys. Rev. C} \textbf{81}, 067302 (2010).

\bibitem[Tsang et~al.(2009)]{PhysRevLett.102.122701}
M.~B. Tsang, Y.~Zhang, P.~Danielewicz, M.~Famiano, Z.~Li, W.~G. Lynch, and
  A.~W. Steiner, \emph{Phys. Rev. Lett.} \textbf{102}, 122701 (2009).

\bibitem[Reinhard et~al.(2006)]{reinhard:014309}
P.-G. Reinhard, M.~Bender, W.~Nazarewicz, and T.~Vertse, \emph{Phys. Rev. C}
  \textbf{73}, 014309 (2006),
  \urlprefix\url{http://link.aps.org/abstract/PRC/v73/e014309}.

\bibitem[Danielewicz and Lee(2012)]{Danielewicz:2012}
P.~Danielewicz, and J.~Lee, Symmetry energy ii: Isobaric analog states (2012),
  in preparation.

\bibitem[Lesinski et~al.(2006)]{Lesinski:2006cu}
T.~Lesinski, K.~Bennaceur, T.~Duguet, and J.~Meyer, \emph{Phys. Rev. C}
  \textbf{74}, 044315 (2006), \eprint{nucl-th/0607065}.

\end{thebibliography}

\end{document}